\documentclass[aps,pre,twocolumn,superscriptaddress,floatfix]{revtex4-2}
\usepackage{amsmath}
\usepackage{amssymb} 
\usepackage{graphicx}
\usepackage{subfigure}
\usepackage{dcolumn}
\usepackage{epstopdf}
\usepackage{braket}
\usepackage{natbib}
\usepackage{amsfonts}

\usepackage[colorlinks, bookmarks=true,breaklinks=true,linkcolor=blue, citecolor=blue, linktocpage=true, urlcolor=blue]{hyperref}

\begin{document}
	\preprint{APS/123-QED}
	\title{System Symmetry and the Classification of Out-of-Time-Ordered Correlator Dynamics in Quantum Chaos}
	\author{Fuxing Chen}
	\affiliation{School of science, Beijing University of Posts and Telecommunications, Beijing 100876, China}
	\author{Ping Fang}
	\email{pingfang@bupt.edu.cn}
	\affiliation{School of science, Beijing University of Posts and Telecommunications, Beijing 100876, China}
	
	\begin{abstract}
		\noindent\textbf{Abstract.}
		The symmetry of chaotic systems plays a pivotal role in determining the
		universality class of spectral statistics and dynamical behaviors, which
		can be described within the framework of random matrix theory.
		Understanding the influence of system symmetry on these behaviors is
		crucial for characterizing universal properties in quantum chaotic
		systems. In this work, we explore the universality of
		out-of-time-ordered correlator (OTOC) dynamics in quantum chaotic
		systems, focusing on the kicked rotor and the kicked Harper model. By
		modulating the periodically kicked potential, we control system symmetry
		to examine its impact on OTOC dynamics and level spacing distributions.
		Our results show that ensemble-averaged OTOC dynamics exhibit distinct
		universal behaviors depending on system symmetry, enabling
		classification through random matrix theory. These distinctions become
		evident after the localization time in localized regimes and emerge at
		specific time scales corresponding to the translational period of the
		Floquet operator in momentum space under quantum resonance conditions.
		Our findings provide a rigorous understanding of the relationship
		between symmetry and quantum chaotic dynamics, contributing to a deeper
		comprehension of universal behaviors in these systems.

	\end{abstract}
	\maketitle
	
	\section{\label{sec:level1}Introduction}

		The initial sensitivity of classical chaotic systems is reflected in the exponential separation of trajectories in phase space, known as the "butterfly effect", typically described by Lyapunov exponents. However, the concept of trajectory in phase space loses its meaning in quantum systems. In this regard, a so-called "quantum butterfly effect" is believed to be detectable through the Out-of-Time-Order Correlator (OTOC)~\cite{COTLER2018318,PhysRevX.7.031047,shenker_black_2014,García-Mata:2023,Hosur2016}. Initially proposed by Larkin and Ovchinnikov in the context of superconductivity theory~\cite{Larkin1969QuasiclassicalMI}, OTOC has recently gained prominence in the field of quantum chaos due to Kitaev's suggestion of using OTOC to describe chaos in many-body systems~\cite{Kitaev2014}. OTOC is employed to characterize the growth of local operators in many-body systems, exhibiting the so-called butterfly velocity~\cite{Roberts2015,PhysRevLett.117.091602}. For single-body chaotic systems, Galitski discovered that the OTOC of kicked rotor system exhibits exponential growth within the Ehrenfest time~\cite{PhysRevLett.118.086801}. Surprisingly, the rate of this growth corresponds to the Lyapunov exponent of classical systems. This suggests that a tool for describing the exponential sensitivity of quantum chaotic systems may have been found. Additionally, OTOC is regarded as an important indicator of quantum information scrambling~\cite{PhysRevA.94.040302,Swingle2018,Landsman2019,PhysRevA.103.062214,PhysRevLett.131.220404}, playing a crucial role in the study of quantum entanglement~\cite{PhysRevX.8.021013,PhysRevResearch.2.013254,10.21468/SciPostPhys.11.4.074}, quantum thermalization~\cite{10.21468/SciPostPhys.11.4.074,PhysRevE.104.034120,PhysRevB.108.L121108}, and black hole information problems. It is worth noting that the subtle relationship between scrambling and chaos remains a controversial topic~\cite{PhysRevLett.124.140602,PhysRevA.104.043308,PhysRevLett.131.180403}.
		
		The kicked rotor model~\cite{IZRAILEV1990299,SANTHANAM20221}, as a typical platform for studying quantum chaos, possesses advantages such as rich physical phenomena, comprehensive theoretical investigations, and controllable experimental manipulation. Recently, the dynamics of OTOC in kicked rotor systems have gained significant attention. For time scales longer than the Ehrenfest time, the exponential decay of one term in the OTOC is believed to imply the mixing process in classical chaotic phase space, with the decay rate determined by Ruelle-Pollicott resonances~\cite{PhysRevLett.121.210601}. Meanwhile, the OTOC exhibits quadratic growth until saturation (caused by localization) in its dynamical behavior~\cite{hamazaki2018operator,PhysRevA.107.062201}. This quadratic growth behavior is also present under the condition of the principal quantum resonance~\cite{e26030229}, where quantum resonance breaks the freeze on OTOC growth caused by localization. Additionally, the scaling behavior of non-Hermitian OTOC~\cite{PhysRevA.107.062201,PhysRevResearch.4.023004}, and the super-exponential growth in mean-field Gross-Pitaevskii models are also intriguing aspects to explore~\cite{PhysRevB.103.184311}.
		
		It is well known that symmetry plays a crucial role in quantum chaos, directly influencing the characteristics of a system's energy level structure. This originates from the establishment of Bohigas-Giannoni-Schmit (BGS) conjecture~\cite{PhysRevLett.52.1}, which bridged the gap between random matrix theory (RMT) and the statistical properties of energy spectra. This led to the universal application of random matrix ensembles in quantum chaotic systems. In Floquet systems, the circular ensemble is considered. In general, circular orthogonal ensemble (COE) is applicable to spinless systems that possess time-reversal (TR) symmetry. It is worth noting that parity-time reversal (PT) symmetric systems with real eigenvalues also fall under the COE class~\cite{Graefe_2015,PhysRevE.80.026213,PhysRevE.96.012154}. 
		Systems without any symmetry are applicable to the circular unitary ensemble (CUE) class. Typically, the level spacing distribution, denoting as $P(s)$, serves as a probe for quantum chaos and system symmetry. The level spacing distribution exhibits level repulsion in non-integrable systems, leading to the Wigner-Dyson distribution. And the system's symmetry determines the Dyson index $\beta$ of the distribution ($\beta=1,2$ corresponds to the COE and CUE classes, respectively). In contrast, integrable systems display a Poisson distribution. It is worth mentioning that for kicked rotor systems, under conditions of localization, the maximal chaos scenario shows a transition from Poisson to Wigner-Dyson distribution~\cite{PhysRevB.31.6852,FEINGOLD1987181,IZRAILEV1987250,CHIRIKOV198877}. The universality of spectral statistics implies that they are independent of specific details of the system. For a long time, researchers have been hoping for such simple and universal laws that govern complex systems.
		
		In addition to the statistical characteristics of level spectra, the dynamical behavior also reflects the features of quantum chaos. One question is whether the dynamical behavior also exhibits universal laws regulated by system symmetry. Taking the kicked rotor as an example, relevant explorations have been conducted in both theory and experiments. This includes the universal laws of energy diffusion under different system symmetries, considering weak localization conditions and quantum resonance conditions theoretically~\cite{PhysRevB.72.045108,PhysRevB.92.235437}. Experimentally, it involves manipulating system symmetry, observing coherent back (forward) scattering and the universal one-parameter scaling law of transport~\cite{Hainaut2018}, and measuring the quantum phase transition time for quantum memory effects of initial states under different system symmetry conditions~\cite{PhysRevLett.121.134101}. These studies reflect the profound influence of system symmetry on dynamical behavior. However, most of these studies focus on the issues of energy diffusion and transport, the impact of system symmetry on the dynamical behavior of OTOC remains a mystery.
		
		In this article, we will demonstrate the OTOC dynamics regulated by system symmetry in the kicked system and explore its universal behavior. Our results show that the exponential growth within the Ehrenfest time and the subsequent quadratic growth of OTOC are not affected by system symmetry. System symmetry exhibits its influence solely until the localization time scale for localization conditions and the metal-to-insulator transition time scale for quantum resonance conditions. We will also demonstrate the scaling behavior of OTOC in different universality classes. Under localization conditions, the localized OTOC of the CUE class exhibit higher saturation values. While in the case of quantum resonance, the situation is reversed, with the OTOC of the COE class showing faster growth after the crossover time. Where this crossover time aligns with the translational period of the Floquet operator in momentum space.
		
		The remainder of the paper is organized as follows. We will introduce the kicked rotor system and analyze the symmetry of the Floquet operator under localization conditions and quantum resonance conditions in Sec.~\ref{sec:level2}. In Sec.~\ref{sec:level3}, we will  introduce the OTOC for pure initial state, with zero angular momentum. And we will present numerical results for the OTOC dynamics under localized and quantum resonance conditions, comparing them with the results of the energy level spacing distribution. Additionally, we have also studied the kicked Harper model to discuss the universality of OTOC. Finally, we will summarize and discuss the work presented in this paper in Sec.~\ref{sec:level4}.

	\section{\label{sec:level2} System and symmetry analysis}
	\subsection{\label{sec:level2-1} Generalized quantum kicked rotor}
		Our work focuses on the OTOC dynamical behavior under different symmetry conditions for quantum chaotic system. As a paradigmatic system for studying quantum chaos with controllable symmetries in experiments, we consider the dimensionless generalized quantum kicked rotors (GQKR). Under periodic amplitude of the kicks $K$, the Hamiltonian of the system is given by:
		\begin{equation}
			\hat{H} =\hat{H_0}+K \cos[\hat{\theta} +a(t)]\sum_{m}\delta(t-m). 
			\label{Hamiltonian}
		\end{equation}
		Here $t, m \in \mathbb{Z}$. $\hat{H_0}$ is considered the free rotation Hamiltonian. Typically, for a rotor, $\hat{H_0} = \frac{{\hbar_{e}}^2\hat{n}^2}{2}$. The $\hbar_e$ plays the role of an effective Planck’s	constant. And there is $\hat{p}=\hbar_e\hat{n}$. Position operator $\hat{\theta}$ and momentum operator $\hat{p}$  satisfy the canonical commutation relation $[\hat{\theta},\hat{p}]=i\hbar_e$. 
		
		Sequence $a(t)$ is utilized for the modulation of the phase of the kicking potential. System will be reduced to the standard kicked rotor when $a(t) = 0$, 
		Furthermore, a method for manipulating the symmetry of controllable systems is provided by quasi-periodic modulation $a(t)$. Prior to this, the Floquet operator of such quasi-periodically modulated systems needs to be taken into account.
		The randomly modulation configuration periodic series $\{a(1), a(2), \cdots, a(N)\}$ be considered as $ a(t) \in [0, 2\pi]$ with $a(t)=a(t+N)$. The evolution operator $\hat{U}(t)$ can be written as: $ \hat{U}(t) =  \hat{F}_t \cdots \hat{F}_2  \hat{F}_1$. Here we consider $N=4$, and the 4-periodically-shifted system's Floquet operator is $\hat{F} =\hat{F}_4 \hat{F}_3 \hat{F}_2 \hat{F}_1$, i.e.,
		\begin{equation}
			\hat{F} = \prod_{j=1}^{4}\hat{F}_j,\quad \hat{F}_j =e^{\frac{-i  \hat{H}_0}{2\hbar_e}}e^{\frac{-i  K \cos[\hat{\theta} +a(j)]}{\hbar_e}}e^{\frac{-i  \hat{H}_0}{2\hbar_e}}.
			\label{PQKR_4}
		\end{equation}
		For a given initial state $\ket{\phi_0}$, 
		the Floquet operator $\hat{F}$ enables the determination of stroboscopic dynamics associated with periodic driving:
		\begin{equation}
			\ket{\phi_{t+4}} = \hat{F} \ket{\phi_t}.
		\end{equation}

	\subsection{\label{sec:level2-2} Symmetry analysis}
		From the perspective of random matrices, the universality class of the circular ensemble to which the Floquet operator belongs is determined by the operator's symmetry. Consider two types of operations: TR operation $\hat{\mathcal{T}}$ and PT operation $\hat{\mathcal{P}}\hat{\mathcal{T}}$. There exists a time reversal transformation and PT transformation,
		\begin{equation}
			\begin{aligned}
				&\hat{\mathcal{T}}\hat{n}\hat{\mathcal{T}}^{-1} = -\hat{n},\:&\: 
				&\hat{\mathcal{P}}\hat{\mathcal{T}}\hat{n}(\hat{\mathcal{P}}\hat{\mathcal{T}})^{-1} = \hat{n},\\ 
				&\hat{\mathcal{T}}\hat{\theta}\hat{\mathcal{T}}^{-1} = \hat{\theta},\:&\:
				&\hat{\mathcal{P}}\hat{\mathcal{T}}\hat{\theta}(\hat{\mathcal{P}}\hat{\mathcal{T}})^{-1} =-\hat{\theta},\\
				&\hat{\mathcal{T}}i\hat{\mathcal{T}}^{-1} = -i,\:&\: &\hat{\mathcal{P}}\hat{\mathcal{T}}i(\hat{\mathcal{P}}\hat{\mathcal{T}})^{-1} = -i.\\
			\end{aligned}
			\label{}
		\end{equation}
		
		For the GQKR system, the symmetry of Floquet operator in Eq.~(\ref{PQKR_4}) can be considered. 
		For the condition $a(j) = a(N+1-j)$, the Floquet operator satisfies the symmetry: $\hat{\mathcal{T}}\hat{F}\hat{\mathcal{T}}^{-1} = \hat{F}^{-1}$. This means the system has TR symmetry. For the condition $a(j) = -a(N+1-j)$, the Floquet operator satisfies the symmetry: $\hat{\mathcal{P}}\hat{\mathcal{T}}\hat{F}(\hat{\mathcal{P}}\hat{\mathcal{T}})^{-1} =\hat{F}^{-1}$ and PT symmetry exists. Floquet operators that meet these two symmetry conditions all follow the COE class.
		However, for random $a(j)$, the Floquet operator lacks any symmetry, and therefore, it belongs to the GUE class.

		For quantum resonances, the effective Planck’s constant is commensurable with $4 \pi$, i.e., $\hbar_e = 4\pi p/q$, the system is in a delocalized state, exhibiting metallic and supermetallic phases, and the system's symmetry needs to be carefully considered. The GQKR ($N=4$) Floquet operator is translationally invariant. The translation operation is considered as:
		\begin{equation}
			\hat{T}_q:\;\hat{n} \rightarrow \hat{n}+4q.
		\end{equation}
		The translation symmetry exists when $[\hat{F},\hat{T}_q]=0$. Note that unlike the standard kicked rotor, the translationally invariant period of GQKR system is $4q$. According to Bloch’s theory, the quasi-energy states $\{\ket{\psi_{\theta_0}}\}$ of the Floquet operator can be expanded by the basis of Bloch states,
		\begin{equation}
			\braket{n | \psi_{\theta_0}} = e^{in\theta_0}\ket{\psi_{n,\theta_0}},
		\end{equation}

		\begin{figure}[!tb]
			\includegraphics[scale=0.36]{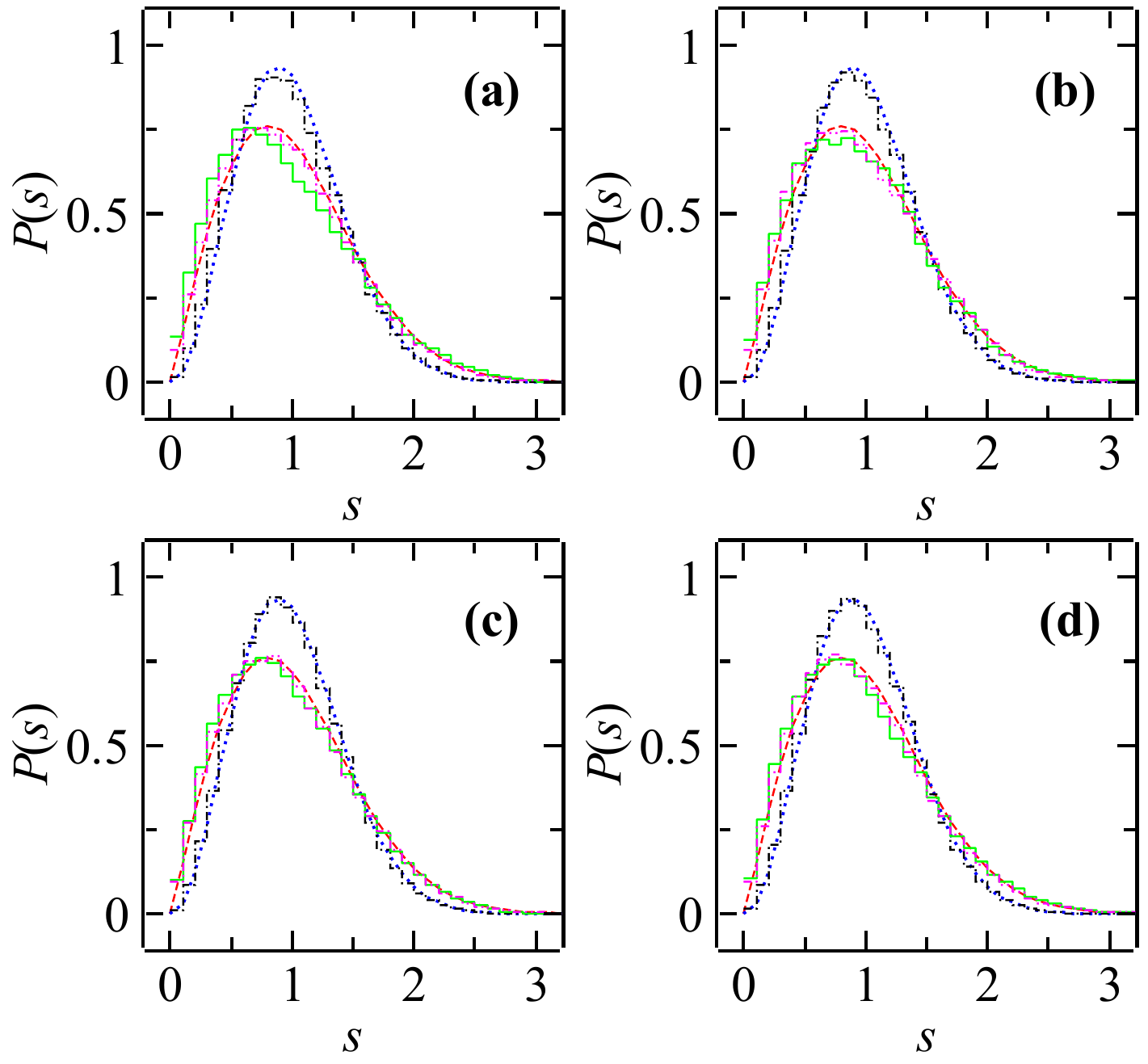}
			\caption{ The level spectrum spacing distribution for parameters: (a) $K=10,\, \hbar_e=0.5$; (b) $K=10,\, \hbar_e=1$; (c) $K=15,\, \hbar_e=0.8$ and (d) $K=20,\, \hbar_e=1$. Here the system with random $a(j)$ (black dot-dash histograms) well compliance CUE class (blue dot curve). And the system with $a(j)=a(N+1-j)$ (green solid histograms) and $a(j)=-a(N+1-j)$ (purple dot-dot-dash histograms) obey COE class (rad dash curve).}\label{fig_local_level}
		\end{figure}

		Here the Bloch number $\theta_0 \in [0,2\pi/4q)$, and Bloch states $\ket{\psi_{n,\theta_0}}=\ket{\psi_{n+4q,\theta_0}}$. The quasi-energy $\{\epsilon(\theta_0)\}$ of Floquet is $\theta_0$ dependent.
		\begin{equation}
			\hat{F}\ket{\psi_{\theta_0}} = e^{-i\epsilon(\theta_0)}\ket{\psi_{\theta_0}}.
			\label{res_Floquet_eig}
		\end{equation}
		Hence, a finite size $\theta_0$-dependent reduced Floquet operator can be represented as:
		\begin{equation}
			\begin{aligned}
				\hat{F}(\theta_0) & = \prod_{j=1}^{N}\hat{F}_j(\theta_0), \\
				\hat{F}_j(\theta_0) & =e^{\frac{-i  \hat{H}_0}{2\hbar_e}}e^{\frac{-i  K \cos[\hat{\theta} +a(j)+\theta_0]}{\hbar_e}}e^{\frac{-i  \hat{H}_0}{2\hbar_e}}.\\
			\end{aligned}
			\label{res_Floquet}
		\end{equation}
		System symmetry needs to be reanalyzed in this instance. In the reduced system, for the condition $a(j) = -a(N+1-j)$, a non-zero Bloch number  will break the original PT symmetry of the Hamiltonian. Consequently, the symmetry $\hat{\mathcal{P}}\hat{\mathcal{T}}\hat{F}(\theta_0)(\hat{\mathcal{P}}\hat{\mathcal{T}})^{-1} =[\hat{F}(\theta_0)]^{-1}$ will be broken for reduced Floquet operator, and the system will belong to the CUE class. However, for the condition $a(j)=a(N+1-j) $, the TR symmetry is preserved and the symmetry of reduced Floquet operator $\hat{\mathcal{T}}\hat{F}(\theta_0)\hat{\mathcal{T}}^{-1} = [\hat{F}(\theta_0)]^{-1}$ still exists. Therefore, the system still belongs to the COE class. Additionally, for random $a(j)$, there is no doubt that the system still belongs to the CUE class.

		\begin{figure}[!t]
			\includegraphics[scale=0.36]{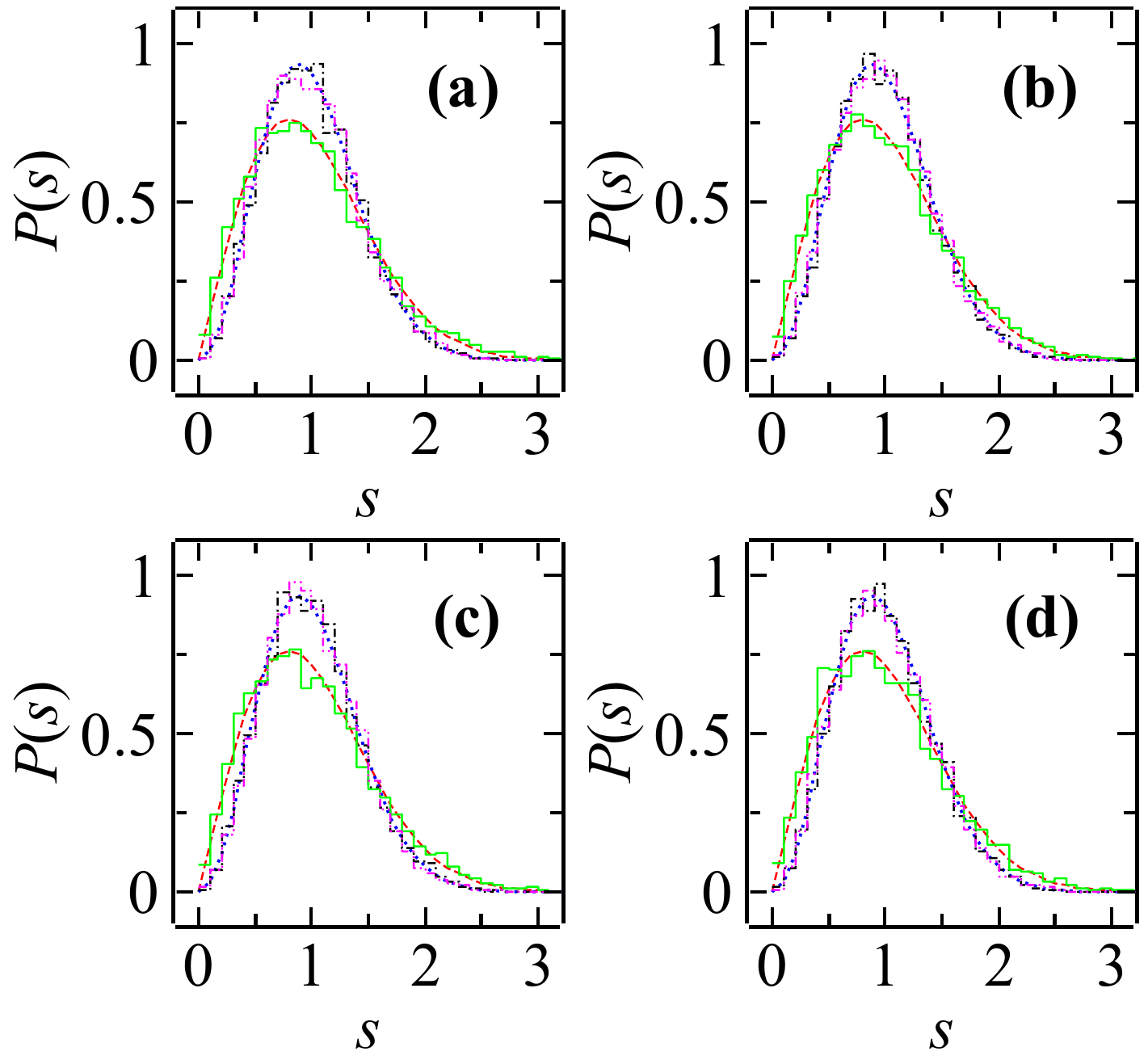}
			\caption{ The level spectrum spacing distribution for parameters: (a) $K=300,\, p/q=4/87$; (b) $K=300,\, p/q=7/87$; (c) $K=300,\, p/q=4/43$ and (d) $K=200,\, p/q=7/87$. Here the system with random $a(j)$ (black dot-dash histograms) and $a(j)=-a(N+1-j)$ (purple dot-dot-dash histograms) well compliance CUE class (blue dot curve). And the system with $a(j)=a(N+1-j)$ (green solid histograms) obey COE class (rad dash curve).}\label{fig_res_level}
		\end{figure}
		
	\section{\label{sec:level3}Level spacing distribution}
		For Floquet systems, the level spacing distribution is determined by the quasi-levels of the Floquet operator. However, the level spacing statistics results must consider the particular context. Consider a Floquet operator in a Hilbert space consisting of a finite number of perturbed states of size $M$. For this, the rotor is now assumed to be defined on a torus instead of on the cylinder. For the dynamical localization, the suppression of chaos is reflected on the scales that exceed the localization length $l_m$. The level spectral spacing distribution exhibits a Poisson distribution. However, to elucidate the influence of symmetry, we focus our spectral analysis on the maximal chaos scenario, i.e., $l_m\gg M$, here $M=10$. As evident in Fig.~\ref{fig_local_level}, in order to ensure statistical significance, 1000 random $a(t)$ are considered. For the Hamiltonian with PT symmetry and TR symmetry, the numerical result of level spacing statistics exhibits a Wigner distribution characteristic of the COE class ($\beta =1$). In contrast, systems lacking symmetry adhere to the Wigner distribution of the CUE class ($\beta=2$).

		For quantum resonances, the size of the Hilbert space is considered to be $M=4q$, taking into account the translationally invariance of the Floquet operator. The parity symmetry of the Hamiltonian is broken in the $\hat{F}(\theta_0)$ unless $\theta_0=0$ or $\pi/M$. To ensure the statistical validity of the level spacing distribution, all values of $\theta_0$ should be considered and randomly sampled. Here 100 random Bloch numbers and $a(t)$ are considered. As shown in Fig.~\ref{fig_res_level}, the level spectrum spacing distribution of four groups with different parameters exhibits good universality. The system with non-symmetry and PT symmetry belongs to the CUE class and the system with TR symmetry belongs to the COE class, as expected.

	\section{\label{sec:level4}The dynamics of OTOC}
	\subsection{\label{sec:level4-1} OTOC of quantum kicked systems for initial localized state }
		The OTOC related to momentum operator can be formulated as follows:
		\begin{equation}
		\begin{aligned}
			 C_p&(t)\\  
			 		=& -\bra{\phi_0}[\hat{p}(t)^\dagger, \hat{p}(0)]^2 \ket{\phi_0}\\
			        =& \bra{\phi_0}\hat{p}(t)^\dagger {\hat{p}(0)}^2 \hat{p}(t)^\dagger \ket{\phi_0}
			       +\bra{\phi_0}\hat{p}(0){[\hat{p}(t)^\dagger]}^2 \hat{p}(0)\ket{\phi_0}\\
			        & -2Re[\bra{\phi_0}\hat{p}(t)^\dagger \hat{p}(0)\hat{p}(t)^\dagger \hat{p}(0)\ket{\phi_0}].\\		        
		\end{aligned}
		\end{equation}
		In the Heisenberg picture, here, $\hat{p}(t)^\dagger =\hat{p}(t) = \hat{U}(t)^\dagger \hat{p} \hat{U}(t) $. For the sake of simplification, we consider the system to be in a pure initial state with zero angular momentum, i.e., $\ket{\phi_0}=\ket{0}$, here $\hat{p}(0)\ket{0}=0\ket{0}$. Hence, the OTOC can be simplified and expressed as follows:
		\begin{equation}
			C_p(t) = \bra{\phi_0}\hat{p}(t) \hat{p}(0)\hat{p}(0)\hat{p}(t) \ket{\phi_0} \equiv I_p(t). 	        
		\end{equation}
		In most cases, assuming the initial state to be $\ket{\phi_0}=\ket{0}$ seems too strict. However, fortunately, for the chaotic systems with initially localized states, $I_p(t)$ dominates the dynamics of OTOC~\cite{hamazaki2018operator}. In the interaction picture, $I_p(t)$ can be written as:
		\begin{equation}
			I_p(t) =  \bra{\phi_t} \hat{p}^2 \ket{\phi_t} \bra{\tilde{\phi}_{0}} \hat{p}^2 \ket{\tilde{\phi}_{0}},   
			\label{OTOC_kin}
		\end{equation}
		where,
		\begin{equation}
			\ket{\tilde{\phi}_{0}} = \hat{U}(t)^\dagger \frac{\hat{p} \ket{\phi_t}}{\sqrt{ \braket{\hat{p} \phi_t|\hat{p} \phi_t}}}.  
		\end{equation}
		It is evident that the first term $\bra{\phi_t} \hat{p}^2 \ket{\phi_t}$ 
		represents the system's kinetic energy at time $t$. And the term $\bra{\tilde{\phi}_{0}} \hat{p}^2 \ket{\tilde{\phi}_{0}}$ 
		illustrates a physical process: Normalize $\ket{\phi_t}$ after applying the momentum operator, then evolve it backwards to the moment at $t=0$ to obtain $\ket{\tilde{\phi}_{0}}$. Finally, calculate the kinetic energy of $\ket{\tilde{\phi}_{0}}$. 
		
		For the GQKR system, the dynamic fluctuations induced by randomly chosen sequence $a(t)$ can be eliminated by ensemble averaging. The average OTOC can be denoted as $\overline{I_p}(t)$

	\subsection{\label{sec:level4-2} OTOC dynamics in localized kicked rotor: short vs. long timescales}
		Under localization conditions, theoretical and experimental investigations reveal a profound distinction in the energy transport dynamics occurring~\cite{PhysRevB.72.045108,Hainaut2018,PhysRevLett.121.134101}, which is dictated by the intrinsic symmetry of the system. In the same vein, it is intriguing to explore the dynamical consequences of these symmetries on OTOC.
		
		Firstly, we consider the OTOC dynamics on a shorter time scale, specifically the Ehrenfest timescale. One function of the OTOC is to measure the quantum analogue of the Lyapunov exponent. Within the Ehrenfest timescale, quantum dynamics correspond to classical dynamics. The classical form of the OTOC is given by:
		\begin{equation}
			C_{p,cl}(t) = {\hbar_{e}}^2 \left\langle \left(\frac{\partial p(t)}{\partial x(0)}\right)^2 \right\rangle,       
		\end{equation}
		here $\langle*\rangle$ represents phase space averaging. The $C_{p,cl}(t)$ can be calculated by considering the classical fully differential relation.
		\begin{equation}
			\begin{aligned}
				&\frac{d p(t)}{d t} \equiv f_1(p,x,t),\\
				&\frac{d x(t)}{d t} \equiv f_2(p,x,t),\\
				&\frac{d}{d t}\left(\frac{\partial p(t)}{\partial x(0)}\right) = \frac{\partial f_1}{\partial p(t)}\frac{\partial p(t)}{\partial x(0)} + \frac{\partial f_1}{\partial x(t)}\frac{\partial x(t)}{\partial x(0)},\\
				&\frac{d}{d t}\left(\frac{\partial x(t)}{\partial x(0)}\right) = \frac{\partial f_2}{\partial p(t)}\frac{\partial p(t)}{\partial x(0)} + \frac{\partial f_2}{\partial x(t)}\frac{\partial x(t)}{\partial x(0)}.\\
			\end{aligned} 	        
		\end{equation}
		Note that $\frac{\partial x(0)}{\partial x(0)}=1$ and $\frac{\partial p(0)}{\partial x(0)}=0$. Hence, for the GQKR system, the classical map is as follows:
		\begin{equation}
			\begin{aligned}
				&p_{t+1} = p_t + K \sin[x_t +a(t)],\\
				&x_{t+1} = x_t + p_{t+1},\\
				&\frac{\partial p(t+1)}{\partial x(0)} = \frac{\partial p(t)}{\partial x(0)} + K \cos[x_t + a(t)]\frac{\partial x(t)}{\partial x(0)},\\
				&\frac{\partial x(t+1)}{\partial x(0)} = \frac{\partial x(t)}{\partial x(0)}+\frac{\partial p(t+1)}{\partial x(0)}.\\
			\end{aligned} 	        
		\end{equation}
		For strongly chaotic conditions $K > 4.5$, the series $\cos[x_t +a(t)]$ can be considered disordered and, more importantly, symmetry independent. The classical OTOC is calculated as:
		\begin{equation}
			\begin{aligned}
			C_{p,cl}(t) &= {\hbar_{e}}^2 \left( \frac{K^2}{2}\right)^t = {\hbar_{e}}^2 e^{\ln(\frac{K^2}{2})t}\\
			&= {\hbar_{e}}^2 e^{2(\ln \frac{K}{2} + \ln \sqrt{2})t}.\\
			\end{aligned}
			\label{OTOC_tE} 	        
		\end{equation}
		\begin{figure}[!tb]
			\includegraphics[scale=0.61]{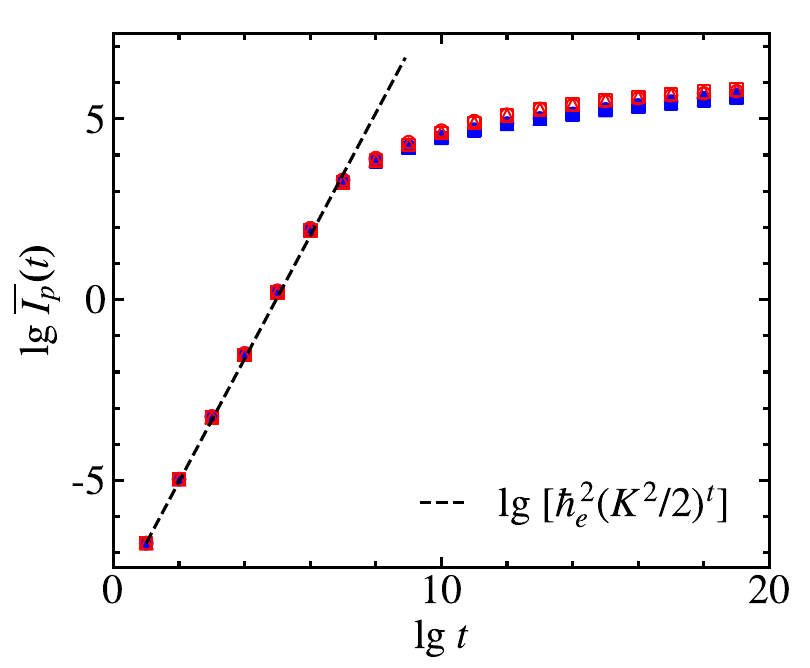}
			\caption{Considering localization ($\hbar_e=2^{-14}$, blue full symbols) and quantum resonances ($\hbar_e=4\pi /205873 \approx 2^{-14}$, red empty symbols) for strongly chaotic conditions $K=10$, $\overline{I_p}(t)$ is the result of averaging $100$ samples with random $a(t)$ for non symmetry [random $a(j)$, square], TR symmetry [$a(j)=-a(N+1-j)$, triangle] and TR symmetry [$a(j)=a(N+1-j)$, circle], respectively. $I_{p}(t)$ is symmetry independent and corresponds to $C_{p,cl}(t)$ in the Ehrenfest time scale, here $t_E \approx 6$. }\label{fig_tE-at}
		\end{figure}
		As shown in Fig.~\ref{fig_tE-at}, in the Ehrenfest time scale, where Ehrenfest time $t_E \approx	\frac{|\ln(\hbar_e)|}{\ln(K/2)}$, $I_{p}(t)$ grows exponentially and corresponds to $C_{p,cl}(t)$ as expected. 
		To ensure that the Ehrenfest time is sufficiently long to observe the short-time exponential growth of the OTOC, we consider the cases of $\hbar = 2^{-14}$ and $\hbar_e = \frac{4\pi}{205873} \approx 2^{-14}$. Although the latter might appear to satisfy a resonance condition, the period of its evolution operator is actually much longer than the theoretical localization length. Consequently, the wave packet remains localized and does not reach an extended state.	Eq.~(\ref{OTOC_tE}) addresses a small debate about the constant differences of $\ln \sqrt{2}$ between the OTOC's growth rate and the Lyapunov exponent $\lambda \approx \ln(K/2)$~\cite{PhysRevLett.118.086801}.

	
		
		
		The OTOC dynamics beyond the $t_E$ timescale are worth discussing. For OTOC dynamics, from Eq.~(\ref{OTOC_kin}) we observe that the OTOC is conceived as a product of two kinetic energies. This implies a potential susceptibility of the OTOC to Anderson localization. Disorder-induced Anderson localization inhibits transport phenomena within the system. For quantum transport, the `break time' $t_b$ emerges as a pivotal concept, representing the temporal scale at which transport dynamics succumb to the pervasive influence of localization. This characteristic timescale exhibits a direct proportionality to the localization length of the underlying system, i.e., $t_b \sim l_m $. Here the localization length $l_m$ is an ambiguous concept. 
		Band random matrix description had given the relationship between localization length and bandwidth $b$~\cite{Jiangbin_1999}: $l_m\propto b$.
		
		\begin{figure}[!tb]
			\includegraphics[scale=0.52]{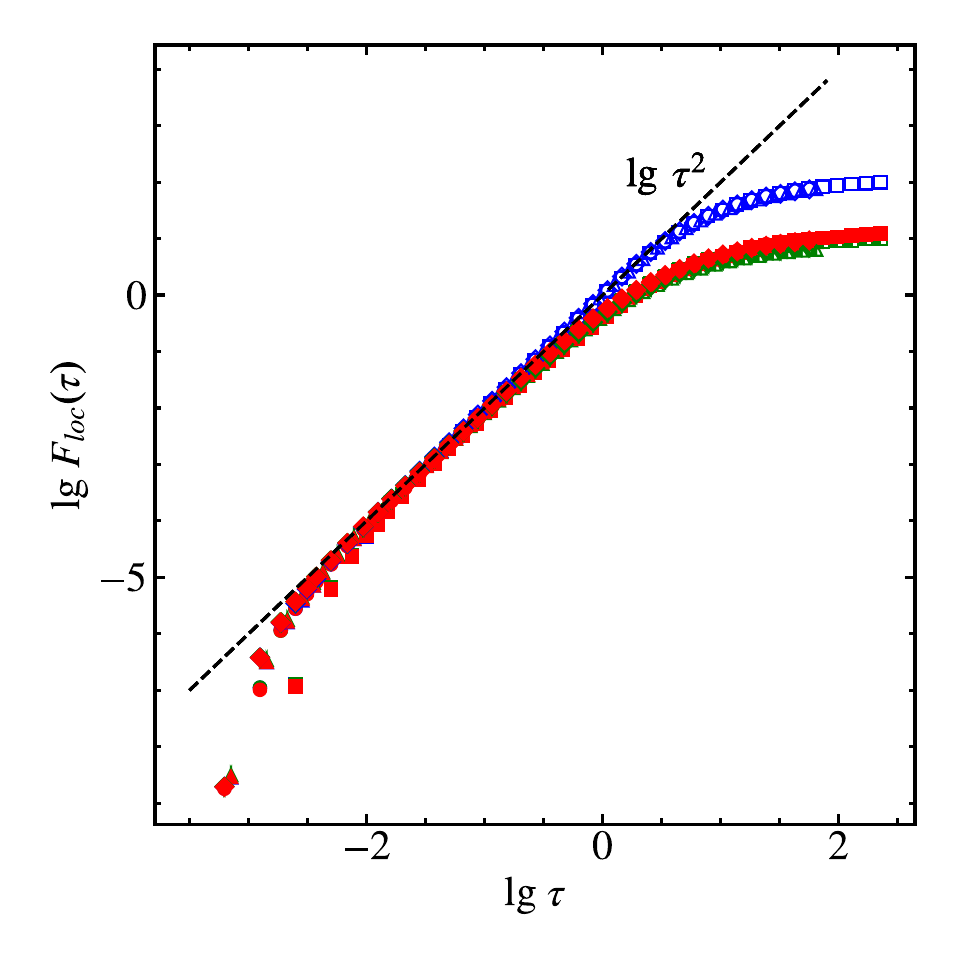}
			\caption{The OTOC for parameters: $K=10,\, \hbar_e=0.5$ (circle); $K=10,\, \hbar_e=1$ (triangle); $K=15,\, \hbar_e=0.8$ (square) and $K=20,\, \hbar_e=1$ (diamond). Here the system with random $a(j)$ (blue empty symbols) obey CUE class. And the system with $a(j)=a(N+1-j)$ (red full symbols) and $a(j)=-a(N+1-j)$ (green half-full symbols) obey COE class.}\label{fig_weak_loc}
		\end{figure}

		For the generalized quantum kicked rotor ($N=4$), 
		the Floquet operator is band matrix in momentum representation and whose bandwidth $b \sim 4K/\hbar_e$. 
		Here, the localization length can be consider to be $l_m \sim 4 D_{cl}/{\hbar_{e}}^2$ for COE class due to bandwidth $b$~\cite{Jiangbin_1999}, where $D_{cl} \approx K^2/4$ denotes the diffusion coefficient. 
		
		In this work, we unveil a scaling OTOC for localization scenarios, 
		\begin{equation}
			F_{loc}(\tau) =\frac{\overline{I_p}(t)}{(2D_{cl})^2 (4D_{cl}/{\hbar_{e}}^2)^2},\qquad  \tau=\frac{t}{4D_{cl}/{\hbar_{e}}^2}.
			\label{scaling_loc}
		\end{equation} 
		As depicted in Fig.~\ref{fig_weak_loc}, the reduced OTOC $F_p(\tau)$ exhibits a dynamical behavior of quadratic growth in $t\gg t_E$. Here the term $(2D_{cl})^2$ can be understood as the growth rate of $\overline{I_p}(t)$ in the quadratic growth regime. Subsequently, it saturates to a constant value due to localization, and $t_b \propto 4D_{cl}/{\hbar_{e}}^2$ serves as the scaling factors. Remarkably, data for different system parameters collapse onto two distinct curves (except for $t < t_E$), and this classification in the dynamics behavior is consistent with the distribution of energy level spacing (as shown in Fig.~\ref{fig_local_level}). This reveals a striking disparity in the dynamics of the OTOC dictated by the system's symmetry (COE or CUE class). 
		
	\subsection{\label{sec:level4-4}OTOC dynamics in kicked rotor for quantum resonances}

		The crossover universality of metal-supermetal transport dynamics has been reported in the kind of chaotic delta "kicked" system~\cite{PhysRevB.92.235437} for quantum resonances, $\hbar_e = 4\pi p/q$. These crossover behaviors are solely determined by the symmetry of the system, that is, whether it belongs to the COE class or the CUE class. The GQKR dynamics crossover is delayed until $t \approx 4q$. The OTOC dynamics, controlled by the system symmetry, are also worth discussing.
		
\begin{figure}[!tb]
			\includegraphics[scale=0.52]{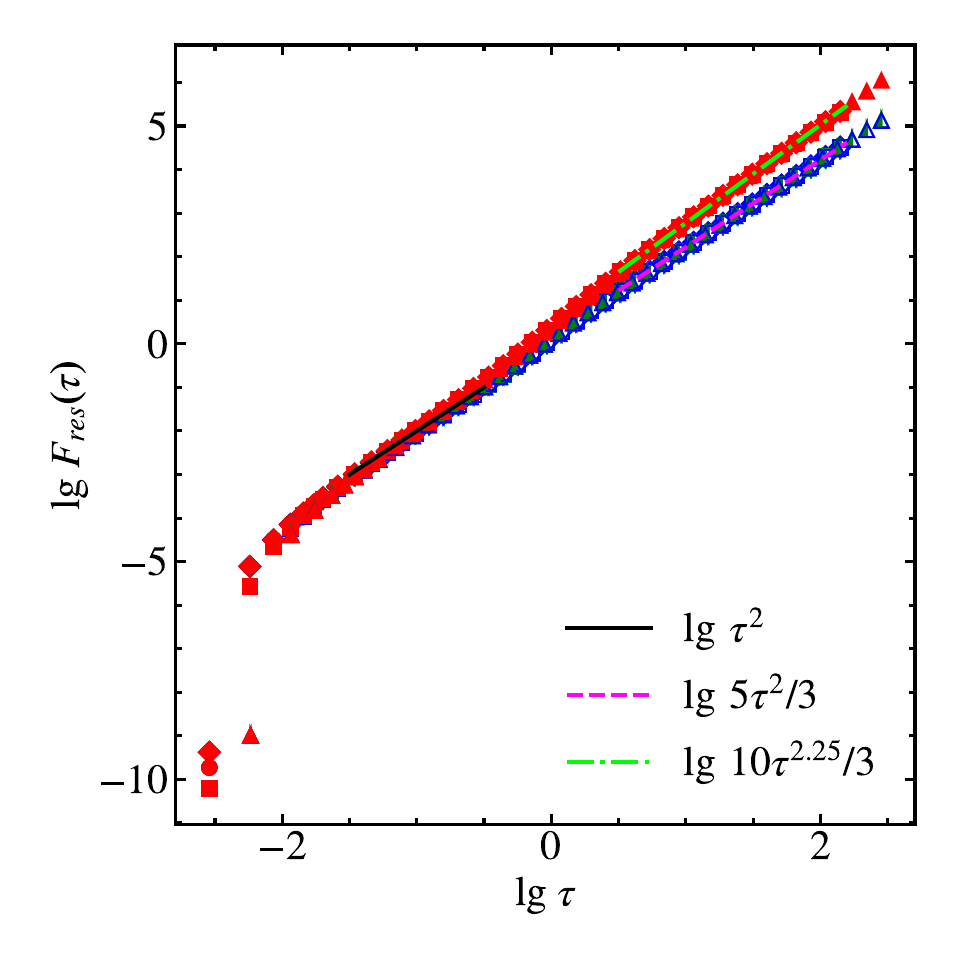}
			\caption{ For quantum resonances, OTOC dynamics is showing for parameters: random $a(j)$ (blue empty symbols), $a(j)=a(N+1-j)$ (red full symbols) and $a(j)=-a(N+1-j)$ (green half-full symbols). Scaling relation shows the university of COE and CUE class, here square, circle, triangle and diamond represents the parameter $(K=300, p/q=4/87), (K=300, p/q=7/87), (K=300, p/q=4/43)$ and $(K=200, p/q=7/87)$, respectively.}\label{fig_res}
		\end{figure}

		Considering the strong chaos condition $K \gg 1$, to ensure localization does not play a role, we require $4q \ll (K/\hbar_e)^2$. For quantum resonance, the scaling relation of $I_p(t)$ and time $t$ is given as:
		\begin{equation}
			F_{res}(\tau) =\frac{\overline{I_p}(t)}{(4q 2D_{cl})^2}=\frac{\overline{I_p}(t)}{( 2q K^2)^2},\qquad \tau=t/4q.
			\label{scaling_res}
		\end{equation}
		And the scaling OTOC dynamics are showing in Fig.~\ref{fig_res}. Given the insensitivity of the strong chaotic quantum resonance condition to the random fluctuations of $a(t)$, we limited our analysis to $20$ sets of randomly generated $a(t)$ samples. 
		It is evident that the OTOC dynamics for $a(j)=-a(N+1-j)$ and random $a(j)$ are the same, and their level spacing distributions are consistent (as shown in Fig.~\ref{fig_res_level}), which implies that we can consider them to belong to the CUE dynamical class.
		Additionally, the systems with TR symmetry, i.e., $a(j)=a(N+1-j)$, has been considered as COE class. It shown a different dynamics behavior from CUE class.  

		For $\tau>t_E/4q$, that is, beyond the Ehrenfest time scale, the data collapse of COE and CUE exhibits universality independent of system details when considering different parameters ($K, \hbar_e, q$).
		It appears that the scaling function $F_{res}(\tau)$ of the CUE class always exhibits quadratic growth. However, it is crucial to pay attention to the specific details. At $\tau \approx 1$, there is a crossover in $F_{res}(\tau)$, transitioning from $\tau^2$ to $5\tau^2/3$. 
		For the COE class, a more pronounced crossover can be clearly observed at approximately $\tau \approx 1$. In this case, the growth behavior of $F_{res}(\tau)$ exhibits a faster power-law trend compared to the CUE class, with the specific form of $10\tau^{2.25}/3$.
		The results indicate that, whether it is the COE or CUE class, the crossover occurs around $t\approx4q$. Therefore, $4q$ can serve as a scaling factor for $F_p(\tau)$. 
		The analysis of the data uncovers the inherent variations in OTOC dynamics across different symmetry classes. This universal classification also aligns with the level spectrum results, as depicted in Fig.~\ref{fig_res_level}. 
		
		\subsection{\label{sec:level4-5} Verifying universality of OTOC in kicked Harper model}
		We have explored the connection between the average OTOC dynamics under conditions of localization and quantum resonance in the kicked rotor model and the universality class of random matrices, and have formulated the corresponding scaling equations. We anticipate that the dynamical characteristics of the OTOC are universal across kicked systems. To validate our conjecture, we here examine the kicked Harper model, which, in contrast to the kicked rotor model Eq.~(\ref{Hamiltonian}), features a modified free Hamiltonian term as: $\hat{H_0} = L \cos(\hbar_{e}\hat{n}+\omega)$. 
		
		Here, localization and quantum resonance are also taken into account. Notably, a non-commensurate $\hbar_e$ with respect to $4\pi$ does not ensure localization. To affirm the system's dynamical localization, we impose the condition $L \gg K$, setting $L = 300000$. The parameter $\omega$ and the sequence $a(t)$ jointly determine the symmetry of the system. Here, $\omega=0, a(j)=a(N+1-j)$ imparts time-reversal symmetry to the system, while $\omega=\pi/2, a(j)=-a(N+1-j)$ imparts PT symmetry to the system. Both belong to the COE class. However, $\omega=\pi/2$ coupled with random $a(j)$ results in a system without any symmetry and is classified under the CUE. 
		
\begin{figure}[!thb]
			\includegraphics[scale=0.58]{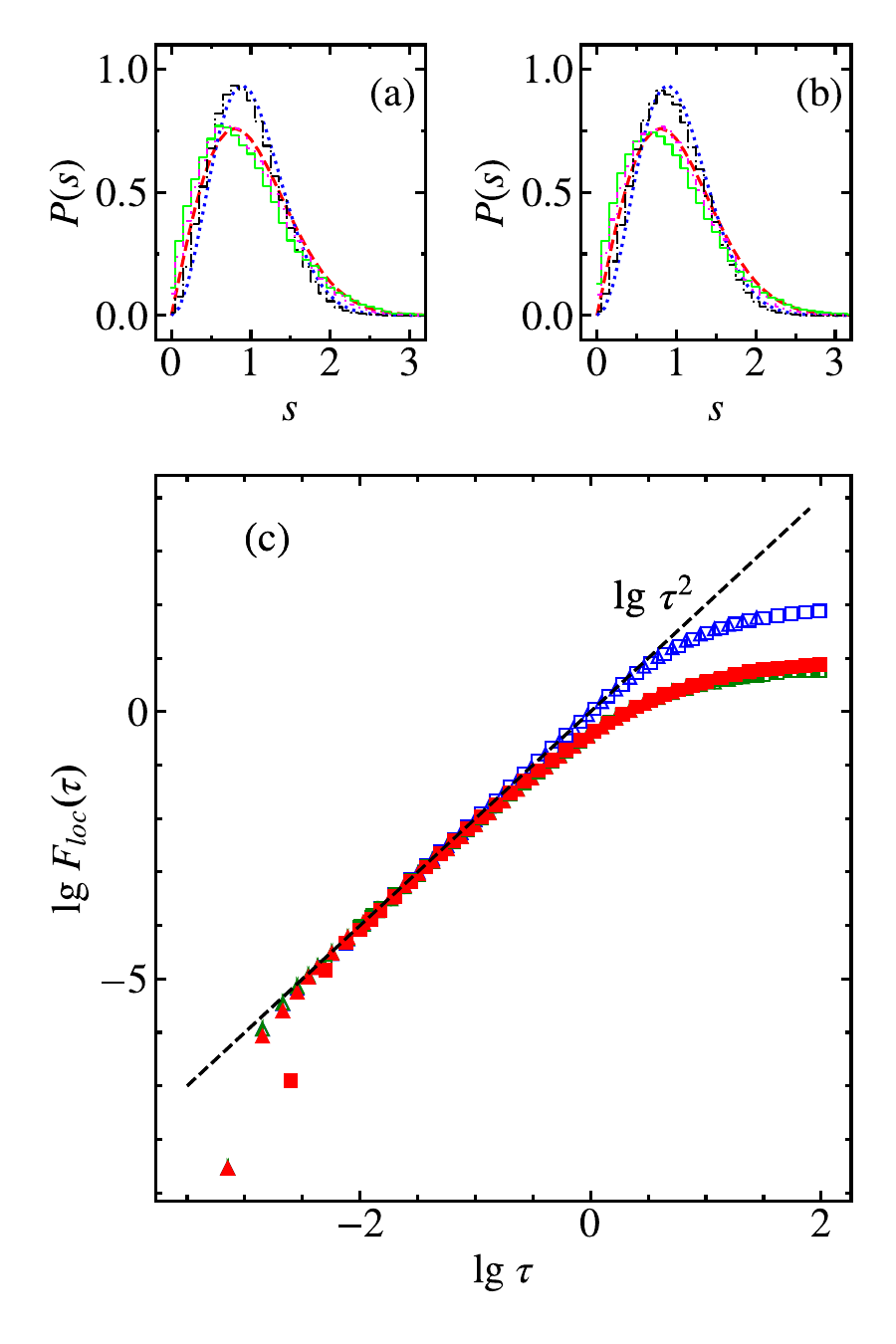}
			\caption{For quantum localization, the level spacing distribution for kicked Harper model with parameters $K=10,\, \hbar_e=1$ (a) and $K=15,\, \hbar_e=0.8$ (b). The corresponding scaling OTOC dynamics (c) are depicted for parameters: $K=10,\, \hbar_e=1$ (triangle); $K=15,\, \hbar_e=0.8$ (square). Here the system with random $a(j)$ (blue empty symbols) obey CUE class. And the system with $a(j)=a(N+1-j)$ (red full symbols) and $a(j)=-a(N+1-j)$ (green half-full symbols) obey COE class.}\label{fig_harper_loc}
		\end{figure}

		As shown in Fig.~\ref{fig_harper_loc}(a) and (b), the level spacing statistics for different parameters and symmetry conditions are displayed under the condition of maximum chaos (i.e., $l_m > M$)". Here, the localization length $l_m$ is considered to be the same as that of the kicked rotor. Additionally, the scaled OTOC Eq.~(\ref{scaling_loc}) demonstrates the dynamical differences between different symmetry conditions in Fig~\ref{fig_harper_loc}(c). It is evident that the scaling relations of OTOCs also apply to the localized kicked Harper model.

\begin{figure}[!tb]
			\includegraphics[scale=0.58]{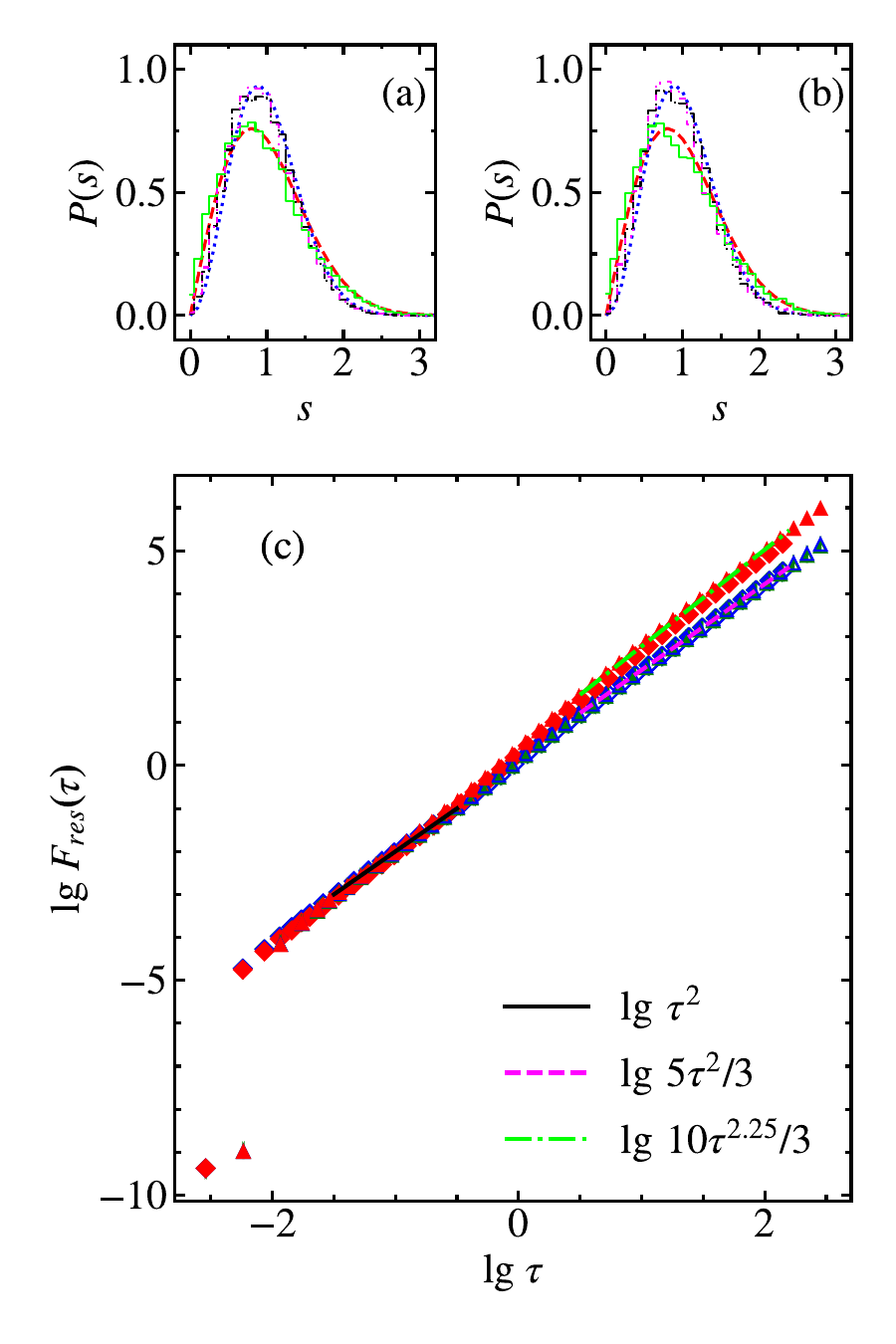}
			\caption{For quantum resonances, the level spacing distribution of kicked Harper model with parameters $(K=200, p/q=7/87)$ (a) and $(K=300, p/q=4/43)$ (b). The corresponding scaling OTOC dynamics (c) are depicted for parameters: random $a(j)$ (blue empty symbols), $a(j)=a(N+1-j)$ (red full symbols) and $a(j)=-a(N+1-j)$ (green half-full symbols). Scaling relation shows the university of COE and CUE class, here triangle and diamond represents the parameter $(K=300, p/q=4/43)$ and $(K=200, p/q=7/87)$, respectively.}\label{fig_harper_res}
		\end{figure}

		For the quantum resonance condition, similarly, $\hbar_e$ is considered to be $\frac{4\pi p}{q}$, and its reduced Floquet operator size $M=4q$. The TR symmetry exists when $\omega=0, a(j)=a(N+1-j)$, corresponding to COE class. However, for the condition $\omega=\pi/2, a(j)=-a(N+1-j)$,as well as for $\omega=\pi/2$ with random $a(j)$, the symmetry is broken, and the system falls into CUE class. In Fig.~\ref{fig_harper_res}(a) and (b), the results of the energy level spacing statistics show the distinction between the symmetry classes of the system, while in Fig.~\ref{fig_harper_res}(c), the crossover behavior of the scaled OTOC Eq.~(\ref{scaling_res}) for different symmetry classes demonstrates consistency with the kicked rotor system.
		
		The consistency of the OTOC dynamics in the kicked Harper model with localization and quantum resonance, and in the kicked rotor, suggests that the dynamical behavior of the OTOC in this class of kicked chaotic systems is universal. Their dynamical differences can be distinguished according to the universality class of random matrices. This universality is determined only by the system's symmetry and allows for the classification of the dynamical differences of OTOC according to the universality class of the random matrix.

	\section{\label{sec:level5}Discussion and summary}
		In this work, we delve into the fascinating model of the GQKR (and kicked Harper) to investigate the dynamic behavior of OTOC under different symmetry conditions. By manipulating the sequence $a(t)$, we can effectively control the symmetry class of the system. In order to showcase the symmetry class of the system, we utilize the distribution of energy level spacings. Notably, the distribution of energy level spacings in the presence of localization conditions is governed by the maximal chaos scenario. Through theoretical analysis and numerical results of the energy level spacing distribution, we are able to identify the symmetry class of the system under both localization and quantum resonance conditions.
		
		One of the most remarkable findings of our study is the scaling form of the average OTOC when considering a pure state with zero angular momentum as the initial state. This scaling form allows us to distinguish between the CUE and the COE classes of OTOC dynamics. This discovery highlights the universal laws governing the OTOC dynamics that are regulated by the system's symmetry.
		
		It is important to note that the influence of system symmetry is not prevalent at all timescales. During the early stages of Lyapunov growth and subsequent quadratic growth, the OTOC dynamics are insensitive to the system's symmetry. However, under localization conditions, the OTOC starts to approach saturation and exhibits symmetry-related dynamic differences at the localization timescale. On the other hand, under quantum resonance conditions, where localization does not occur, at the timescale of $t\approx4q$ (for the GQKR model with $N=4$), the OTOC demonstrates different universal dynamic crossover behaviors influenced by the system's symmetry. For CUE class, the OTOC almost maintains a quadratic growth pattern. However, the OTOC of the COE class exhibits a power-law crossover from an exponent of $2$ to $2.25$. This faster power law is obviously different from the second growth with the principal quantum resonance condition\cite{e26030229}, even though they both have time reversal symmetry.

		The intricate interplay between symmetry conditions and OTOC dynamics in the GQKR model opens up new avenues for understanding the fundamental principles underlying quantum systems. By manipulating the symmetry class of the system, we can gain valuable insights into the behavior of OTOC and its connection to the underlying symmetries. This research not only contributes to the theoretical understanding of quantum dynamics but also paves the way for potential applications in various fields, including quantum information processing and quantum computing.

	\section{ACKNOWLEDGMENTS}
		This research was supported by the Fundamental Research Funds for the Central Universities with contract number 2024ZCJH13,the National Natural Science Foundation of China (Grant No. 12005024), the Fundamental Research Funds for the Central Universities with contract number 2023ZCJH02, and the High-performance Computing Platform of Beijing University of Posts and Telecommunications.
	\bibliography{Ref}
\end{document}